\documentclass[fleqn,usenatbib]{mnras}

\usepackage{newtxtext,newtxmath}

\usepackage[T1]{fontenc}

\DeclareRobustCommand{\VAN}[3]{#2}
\let\VANthebibliography\thebibliography
\def\thebibliography{\DeclareRobustCommand{\VAN}[3]{##3}\VANthebibliography}

\usepackage{graphicx}	
\usepackage{amsmath}	
\usepackage{multicol}
\usepackage{subcaption}
\usepackage{comment}

\title[Magnetic turbulence in Tycho's SNR]{Investigating the nature of magnetic turbulence in Tycho's SNR using X-ray observation}

\author[Aravindaraj et al.]{
Ashwin Aravindaraj,$^{1}$\thanks{E-mail: ashwin@gapp.nthu.edu.tw}
Hui Zhu,$^{2}$\thanks{E-mail: zhuhui@bao.ac.cn}
Adrien Picquenot$^{3}$
and Hsiang-Kuang Chang$^{1,4,5,6}$\thanks{E-mail: hkchang@mx.nthu.edu.tw}
\\
$^{1}$Institute of Astronomy, National Tsing Hua University, Hsinchu 300044, Taiwan\\
$^{2}$State Key Laboratory of Radio Astronomy and Technology, National Astronomical Observatories, CAS, Beijing, 100012, China\\
$^{3}$Department of Physics \& Astronomy, Louisiana State University, 202 Nicholson Hall, Baton Rouge, LA 70803, USA.\\
$^{4}$Department of Physics, National Tsing Hua University, Hsinchu, Taiwan\\
$^{5}$Institute of Space Engineering, National Tsing Hua University, Hsinchu 300044, Taiwan\\
$^{6}$Center for Theory-Computation-Data Science Research (CTCD), National Tsing Hua University, Hsinchu 300044, Taiwan
}

\date{Accepted XXX. Received YYY; in original form ZZZ}

\pubyear{\the\year{}}

\begin{document}
\label{firstpage}
\pagerange{\pageref{firstpage}--\pageref{lastpage}}
\maketitle

\newcommand{\adrien}[1]{\kibitz{green}      {[AP: #1]}}

\begin{abstract}
Supernova remnants (SNRs) are widely regarded as the primary sources of Galactic cosmic ray acceleration and the particles are energized at the shock front of SNRs through the diffusive shock acceleration (DSA) mechanism, gaining energy by repeatedly crossing the shock. Magnetic turbulence plays a crucial role in scattering these particles back and forth, making the investigation of turbulence essential to understanding the acceleration process. In this study, we apply the two-point correlation method to derive the magnetic energy spectrum from the non-thermal X-ray flux image of Tycho’s SNR. This image was obtained using Poissonian-based Generalized Morphological Component Analysis (pGMCA). The turbulence length scale observed in the resulting spectrum is shorter than the turbulence scale derived from previous studies of radio observations of Tycho’s SNR. Additionally, we use the X-ray rim thickness measured across Tycho which is an indicator of magnetic field strength to estimate the magnetic energy spectrum. The turbulence scale inferred from this method is consistent with that obtained from the non-thermal X-ray flux analysis.
\end{abstract}




\section{Introduction}
Supernova remnants (SNRs) are considered to be one of the sources in our Galaxy which are capable of accelerating cosmic ray particles up to PeV energies. The existence of thin X-ray rim around the forward shock of young SNR and the thickness of the X-ray rim reveal that the magnetic field present in the downstream of the shock are amplified up to $100\mu G$ above the mean interstellar magnetic field \citep[][]{2005ApJ...621..793B, 2003ApJ...584..758V, 2006A&A...453..387P}. The existence of this amplified strong magnetic field indicates that the particles are accelerated up to very high energy by the diffusive shock acceleration (DSA) model \citep[][]{1978ApJ...221L..29B, 1978MNRAS.182..147B} at the shock front through repeated crossings between the downstream and upstream of the shock front . Magnetic turbulence plays a major role in scattering the particles back and forth, and thus also in accelerating particles to knee energies of cosmic rays. Tycho's SNR, one of the historic remnants that was discovered by Tycho Brahe in 1572 AD, is observed well in all wavelengths and is studied extensively. VERITAS reported the detection up to 10 TeV \citep[][]{2011ApJ...730L..20A}, indicating that the particles are accelerated up to very high energy. 

A two-point correlation function is widely used to calculate magnetic turbulence scale in the interstellar medium (ISM) and \citet{2012ApJ...747....5L,2016ApJ...818..178L} gave a mathematical description of how two-point correlation gives details about magnetic turbulence scale in those systems. Cosmic-ray particles resonantly interact with turbulence on scales comparable to their gyro-radius. As a result, the scattering frequency depends on both the turbulence scale and the slope of the magnetic energy spectrum. 

\citet{2006A&A...453..387P} showed that the diffusion coefficient in SNRs follows a power-law form: $D(E)\propto E^{1-m}$, where $m$ is the slope of the magnetic energy spectrum. So obtaining the slope of the magnetic energy spectrum is important in clarifying the transport mechanism near the shock-front, which is responsible for accelerating the particles up to very high energy.
\citet{2018MNRAS.480.2200S} calculated the two-point correlation function for Tycho's SNR using radio synchrotron intensity and found a turbulence scale developed up to 126.5" near the shock-front and the magnetic spectrum follows a Kolmogorov-like scaling \citep[][]{1941DoSSR..30..301K} with a scale of 2/3. 
\citet{2006A&A...453..387P} showed that for $m\approx$2/3, the energy of protons reaches knee energy under some compression models for Tycho, Kepler and Cas A SNRs. 
Electrons emitting radio waves are long-lived and usually trace the long-range turbulence present in the system. 
Since the electrons capable of emitting X-ray photons are usually short-lived and are usually near the active shock acceleration sites, these X-ray photons could be used to trace the magnetic turbulence in the plasma near the active acceleration sites where particles are accelerated up to very high energy. 

One difficulty in using the X-ray image is that the thermal emission is dominant in the (1.0 - 4.0 keV) energy range. 
We used the blind source separation algorithm like poissonian based generalized morphological component analysis (pGMCA), which is developed for X-ray astronomy based on poissonian count statistics, to separate the thermal emission from the image and used the non-thermal component in our two-point correlation analysis. 
Since the synchrotron X-ray flux image is dependent on the number density of the relativistic electrons and the magnetic field component perpendicular to the line of sight, the derived magnetic energy spectrum will be dependent on the density of electrons. 
We therefore used the X-ray rim thickness, which is directly dependent on the magnetic field strength alone, to produce the magnetic energy spectrum, which would relate to the turbulence of magnetic field alone and we compare it with the results from our X-ray flux image.

In Section 2, we describe the data analysis method and present the magnetic energy spectrum derived from the non-thermal X-ray flux image. In Section 3, we calculate the X-ray rim thickness and use it to independently estimate the magnetic energy spectrum. Finally, in Section 4, we discuss our results and their implications.
\section{Data Analysis}
To study magnetic turbulence in Tycho’s SNR, high spatial resolution is essential for resolving the turbulence scales present in the system. Therefore, we use Chandra data, which offers superior spatial resolution in the X-ray band. In this section, we describe the data selection process and the application of the pGMCA algorithm to obtain the non-thermal X-ray image. We then outline the two-point correlation method and the analysis steps used to derive the magnetic energy spectrum.

\subsection{Observation}
We have used archival Chandra ACIS-I observation of Tycho between 2009 April 11 and May 3 (PI: J. Hughes; ObsIDs: 10093–10097, 10902–10904, 10906) with a total exposure time of 734 Ks. We have used CIAO to produce the level 2 event file. 

\subsection{pGMCA}
\label{pgmca}

To retrieve an accurate map of the non-thermal emission, we used a blind source separation method based on the General Morphological Components Analysis \citep[GMCA, see][]{bobin15}, first introduced for X-ray observations by \cite{picquenot:hal-02160434}. An updated version, the pGMCA, was developed to take into account Poisson statistics in \cite{9215040}. The pGMCA algorithm can disentangle spectrally and spatially mixed components from an X-ray data cube of the form $(x,y,E)$ with no prior physical and instrumental information by focusing on the morphological diversity of the components to separate.

\begin{figure}
    \includegraphics[width=0.45\textwidth]{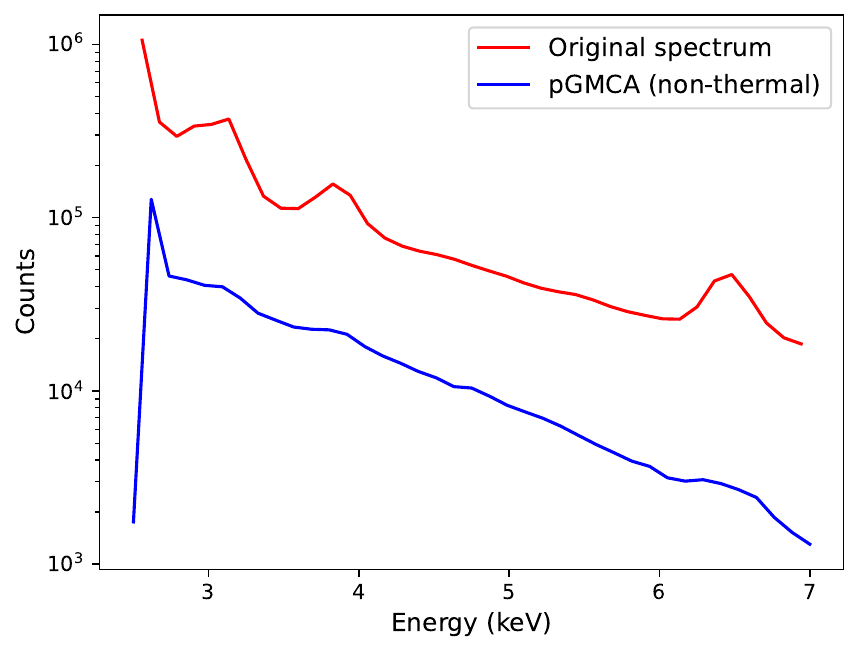}
    \caption{Comparison of the non-thermal spectrum obtained from the pGMCA with the original spectrum derived from the data}
    \label{fig:energy_Count}
\end{figure}

In order to increase the statistics in each pixel, we rebinned spectrally the stacked Chandra observations to $116.8$ eV energy bins but kept the native $0.5$" spatial bin size. We then applied the pGMCA algorithm to the $2.5$ - $7$ keV and $2.5$ - $4$ keV energy ranges, and divided the non-thermal image thus obtained by the exposure map in order to produce flux images. The non-thermal spectrum obtained from the pGMCA is shown in Fig.\ref{fig:energy_Count}. We used the $2.5$ - $7$ keV flux image in the two-point correlation analysis to analyze the magnetic turbulence in these regions.

\subsection{Two-point correlation}
Two-point correlation is a tool to measure the turbulence in the system and the two point correlation of synchrotron intensity from ISM is used to determine the magnetic turbulence. Two-point correlation studies have been applied to radio images of Tycho’s SNR \citep[][]{2018MNRAS.480.2200S} and Cas A \citep[][]{2020MNRAS.498.1093V}. A detailed mathematical description of the method can be found in the appendix of \citet{2018MNRAS.480.2200S}.
As stated in \citet{2018MNRAS.480.2200S} we have selected the points along the concentric circle to neglect the projection effects in the calculation. 
To analyze the nature of magnetic turbulence present in the Tycho SNR, we employed the two-point correlation of the X-ray flux image,
\begin{equation}
    C^{(2)}(\lambda) = \frac{\int F(X)F(X')\mathrm{d^2X}}{\int \mathrm{d^2X}}\,\,\, ,
\end{equation}
where $F(X)$ is the X-ray photon flux at position $X (x,y)$ and $\lambda =X-X'$ is the spatial separation vector between two points.
To remove the contribution of thermal flux from the image we have employed the poissonian based generalized morphological component analysis (pGMCA) which is explained in section \ref{pgmca}. The magnetic energy spectrum is characterized by $|C^2(\lambda)-C^2(\lambda_{\rm min})|$, which removes the contribution from the mean magnetic field (see the appendix of \citet{2018MNRAS.480.2200S} for details). To obtain a dimensionless quantity and facilitate comparison between different regions, the spectrum was normalized by its maximum value: $\frac{|C
^2(\lambda)-C^2(\lambda_{\rm min})|}{max(|C
^2(\lambda)-C^2(\lambda_{\rm min})|)}$

\subsection{Magnetic energy spectrum using non-thermal flux image}
\begin{figure}    
\includegraphics[width=0.45\textwidth]{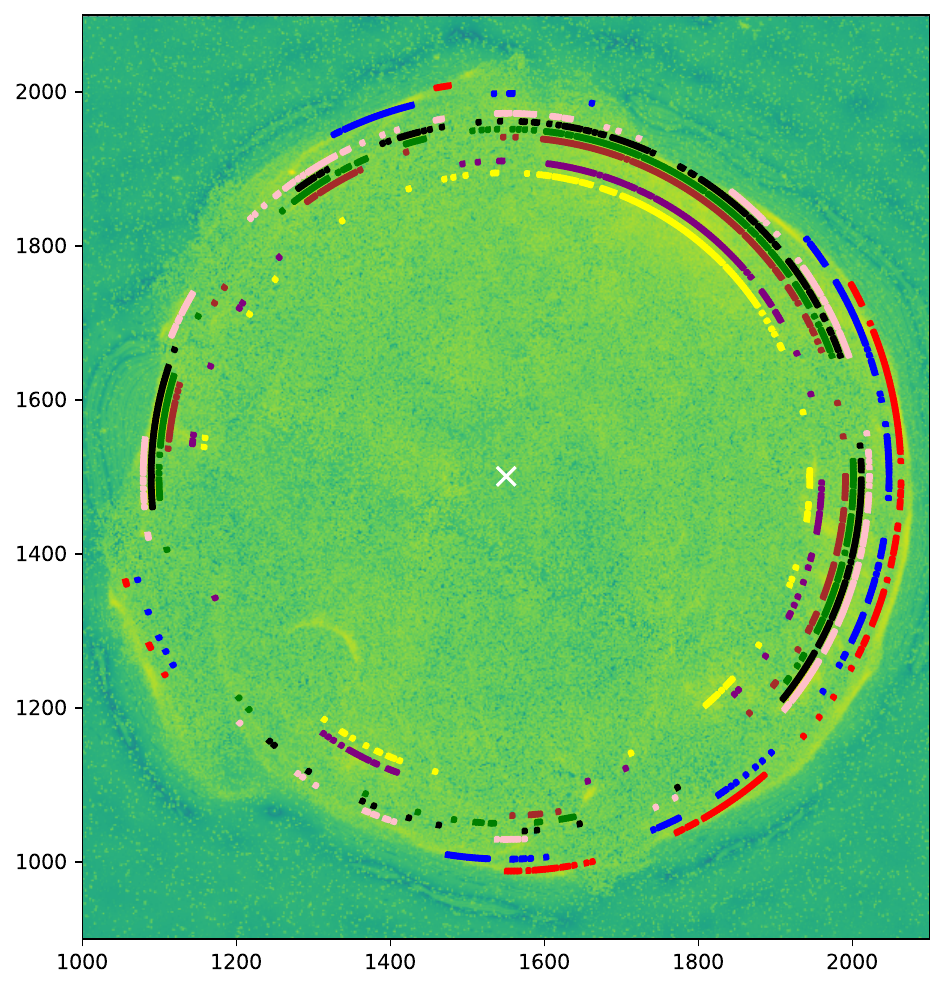}
    \caption{X-ray photon flux image (2.5-7.0 keV) of Tycho's SNR. The different colored concentric circles show the region selected for the correlation function computation. The color of the concentric circles represents different radii as the following: red-$R_{\rm SNR}$, blue-$0.97R_{\rm SNR}$, pink-$0.94R_{\rm SNR}$, black-$0.90R_{\rm SNR}$, green-$0.88R_{\rm SNR}$, brown-$0.85R_{\rm SNR}$, purple-$0.82R_{\rm SNR}$, yellow-$0.79R_{\rm SNR}$. The cross-mark in the image indicates the centre of the remnant.}
    \label{con_circle}
\end{figure}

We analyzed the non-thermal flux image in the energy range of 2.5-7.0 keV using the two-point function for various concentric circles of radius $R_{\rm SNR}$, $0.97 R_{\rm SNR}$, $0.94 R_{\rm SNR}$, $0.90 R_{\rm SNR}$, $0.88 R_{\rm SNR}$, $0.85 R_{\rm SNR}$, $0.82 R_{\rm SNR}$, and $0.79 R_{\rm SNR}$, where $R_{\rm SNR}$=253 arcsec as shown in Fig.\ref{con_circle} and the centre of the concentric circle is set at (R.A.=$0^{\rm h}25^{\rm m}19^{\rm s}.9$, 
Dec=$+64^{\circ}08'18''.2$) as mentioned in \citet{Ruiz_Lapuente_2004}. The thickness of each concentric circle is 0.0091 $R_{\rm SNR}$ and the minimum separation ($\lambda_{\rm min}$) is 1.968 arcsec. The PSF of chandra for this source is about 1 arcsec and the minimum separation was chosen so that its larger than the PSF.
The noise level of the image, (SD) is $\sigma=5\times10^{-10}$ photons cm$^{-2}$ s$^{-1}$.
To exclude regions dominated by noise, each concentric circle was divided into segments with a thickness of $0.0091R_{\rm SNR}$ and a width of 1.968 arcsec. If the mean flux within a given segment was less than $3\sigma$, that segment was excluded from the two-point correlation calculation. The upper limit of the correlation function was estimated by assigning a value of $3\sigma$ to each individual pixel with flux below $3\sigma$, while the lower limit was obtained by assigning a value of 0 to those pixels.


\begin{figure*}
  \begin{subfigure}{\columnwidth}
  \includegraphics[width=\textwidth]{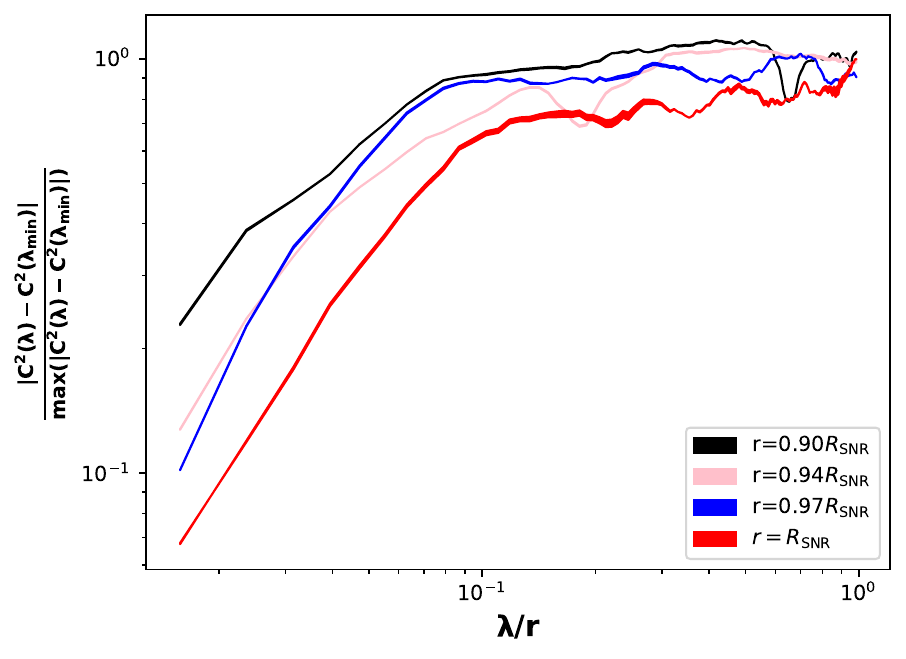}
  \end{subfigure}
  \hfill
  \begin{subfigure}{\columnwidth}
  \includegraphics[width=\textwidth]{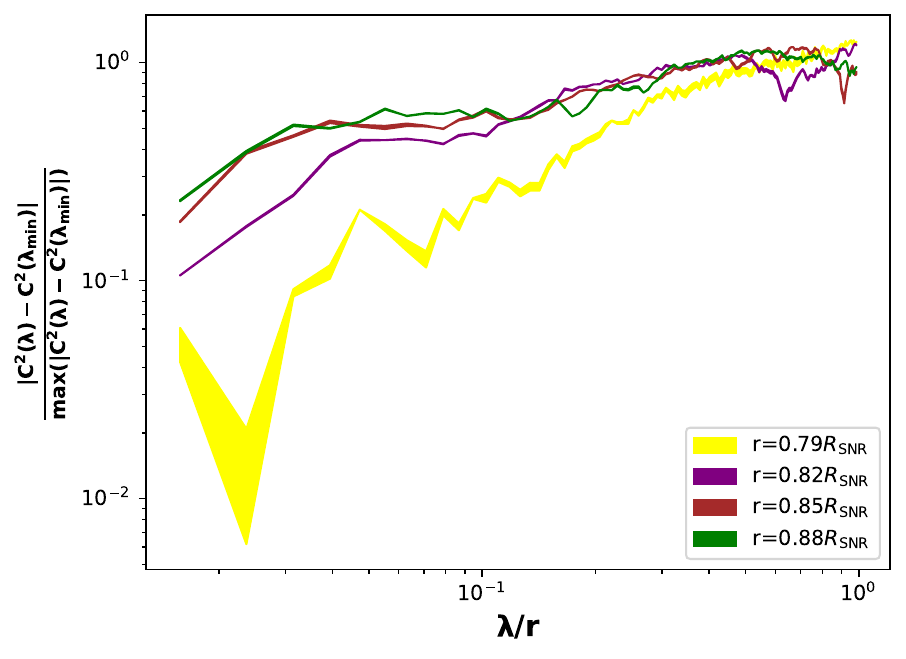} 
  \end{subfigure} 
  \caption{Magnetic energy spectrum for X-ray photon flux image (2.5-7.0 keV) of Tycho's SNR. Thickness of the line corresponds to the value range of the correlation function as explained in the main text. The left panel shows the correlation functions for circles of  $r>0.90R_{\rm SNR}$ and the right panel shows the others. }
  \label{Two_point_flux}
\end{figure*}

To characterize the nature of the magnetic turbulence, we plotted $\frac{|C
^2(\lambda)-C^2(\lambda_{\rm min})|}{max(|C
^2(\lambda)-C^2(\lambda_{\rm min})|)}$ as a function of the corresponding separation $\lambda$ for different concentric circles. For $r \geq 0.9R_{\rm SNR}$, the turbulence is developed only up to $\lambda/r \leq 0.1$, as shown in the left panel of Fig.~\ref{Two_point_flux}, and follows a power-law scaling significantly steeper than 0.6 (see Table~\ref{pl_twopoint}). This differs from the radio synchrotron study of \citet{2018MNRAS.480.2200S}, in which turbulence extended up to $\lambda/r \leq 0.5$ with a scaling of $\lambda^{0.66}$. In that work, the reduction of the turbulence scale from 0.5 to 0.2 toward the inner region ($r \leq 0.9R_{\rm SNR}$) was attributed to the onset of Rayleigh--Taylor instability (RTI), with the corresponding contact discontinuity located at $r = 0.9R_{\rm SNR}$ \citep[][]{2005ApJ...634..376W}. Similarly, in the X-ray image, the turbulence spectrum derived from concentric circles with $r \leq 0.9R_{\rm SNR}$ becomes highly irregular and does not show a distinct turbulence scale. It should be noted that the derived spectrum is affected by geometrical projection and shock irregularity, given that the non-thermal X-ray emission is confined to a thin shock-front rim.
\begin{table}
	\centering
	\caption{The slope of the magnetic energy spectrum for concentric circles with radii $r\geq0.9R_{\rm SNR}$ in the X-ray photon flux image (2.5-7.0 keV) of Tycho's SNR}
	\label{pl_twopoint}
	\begin{tabular}{lr} 
		\hline
		radius & slope (m)\\
		\hline
		${R_{\rm SNR}}$ & 
            $1.29\pm0.08$\\
            ${0.97R_{\rm SNR}}$ & $1.05\pm0.05$\\
            ${0.94R_{\rm SNR}}$ & $1.08\pm0.07$\\
            ${0.90R_{\rm SNR}}$ & $1.16\pm0.06$\\
        
		\hline
	\end{tabular}
\end{table}

\section{X-ray rim thickness}
The two-point correlation function of the X-ray flux depends on both the magnetic field strength and the density of relativistic electrons. As a result, the inferred turbulence may be influenced by both factors. To isolate the magnetic contribution, we need a parameter that depends solely on the magnetic field strength.

The thin X-ray rim observed at the shock front of Tycho's SNR, as well as in other young supernova remnants, indicates the presence of synchrotron X-ray emission \citep[][]{2005ApJ...621..793B}. To explain these narrow rims, two models have been proposed: the energy-loss model and the magnetic damping model \citep[][]{2005pohl}. In an effort to distinguish between the two in Tycho’s SNR, \citet{2015ApJ...812..101T} examined how the rim thickness varies with energy. However, their results showed that neither model could uniquely account for the observed rim thickness.The damping length inferred from the joint modeling of X-ray and radio data was larger than the X-ray rim thickness, indicating that the electron energy loss is important in both models. Given this, and for the sake of simplicity, we assume in this study that the X-ray rim thickness can serve as an indicator for magnetic field strength
\citep[][]{2014ApJ...790...85R}. 

In this study, we measure the X-ray rim thickness around the shock front and use it as a proxy for the magnetic field strength. We apply a two-point correlation analysis to the derived rim thickness. This new approach helps reduce the contribution of density fluctuations in the inferred magnetic energy spectrum. To further minimize energy-dependent effects, the analysis is restricted to a narrow energy band (2.5–4.0 keV) 

\subsection{Magnetic energy spectrum using X-ray rim thickness}
To measure the thickness of the X-ray filaments, we fitted the filament profile with a piecewise two-exponential model as given in \citet{2015ApJ...812..101T}:
\begin{equation}
h(r) =
\begin{cases} 
A_{\rm d} \exp \left( \frac{r - r_0}{w_{\rm d}} \right) + C_{\rm d}, & r < r_0 \\
A_{\rm u} \exp \left( \frac{r_0 - r}{w_{\rm u}} \right) + C_{\rm u}, & r \geq r_0
\end{cases}
\label{eq2}
\end{equation}
where $h(r)$ is the profile height and $r$ is the radial distance from the centre. $A_{\rm d}$ and $A_{\rm u}$ are the normalization factors in the upstream and downstream regions, respectively; $W_{\rm d}$ and $W_{\rm u}$ are the corresponding widths; and $C_{\rm d}$ and $C_{\rm u}$ are the background levels in the upstream and downstream, respectively. To ensure continuity at $r = r_0$, we use $A_{\rm d} = A_{\rm u} + (C_{\rm u} - C_{\rm d})$.

To reduce the energy dependence, we have used the narrow energy range (2.5-4 keV) image obtained using pGMCA to measure the thickness of the X-ray rim. \citet{2023A&A...672A..57P} have used the non-thermal image of Cassiopeia A obtained from pGMCA to calculate the X-ray rim thickness and we followed the same method to calculate FWHM from non-thermal X-ray image (2.5-4.0 keV) of Tycho to calculate the FWHM of upstream and downstream filament profile. The X-ray rim thickness is calculated by taking the average of these two FWHM. 

\begin{figure}
    \includegraphics[width=0.45\textwidth]{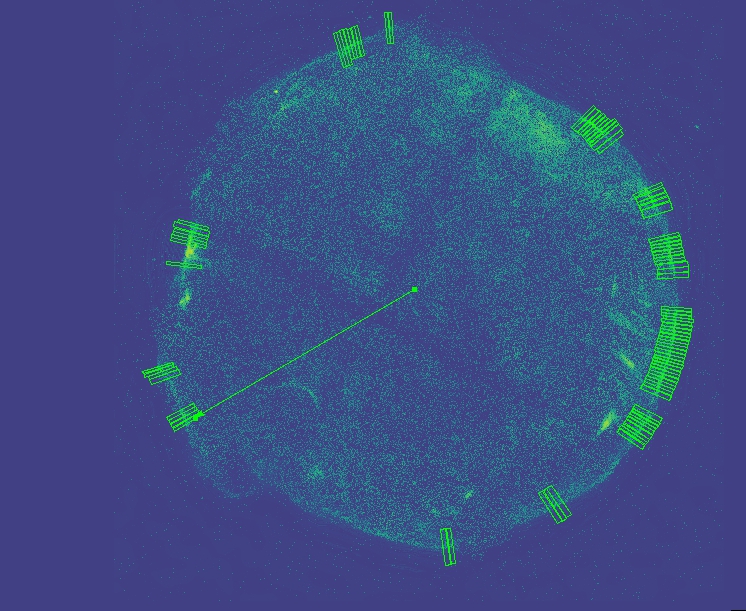}
    \caption{Regions selected to measure the rim thickness in the X-ray image (2.5-4.0 keV) of Tycho's SNR}
    \label{region_selection}
\end{figure}

\begin{figure*}
  \centering
  \begin{subfigure}{0.32\textwidth}
    \includegraphics[width=\textwidth]{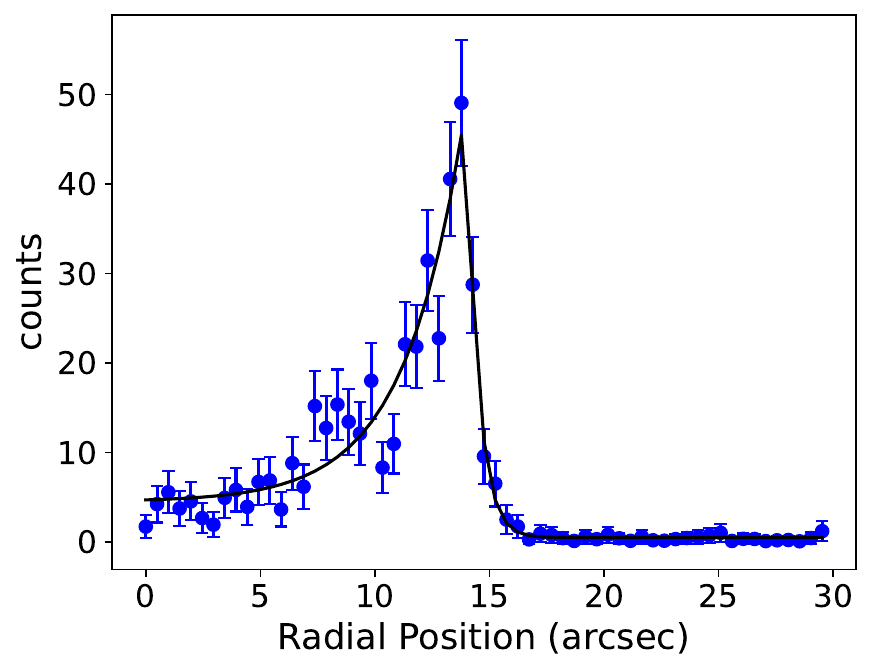}
  \end{subfigure}
  \hfill
  \begin{subfigure}{0.32\textwidth}
    \includegraphics[width=\textwidth]{figure/4.5-7_2.pdf}
  \end{subfigure}
  \hfill
  \begin{subfigure}{0.32\textwidth}
    \includegraphics[width=\textwidth]{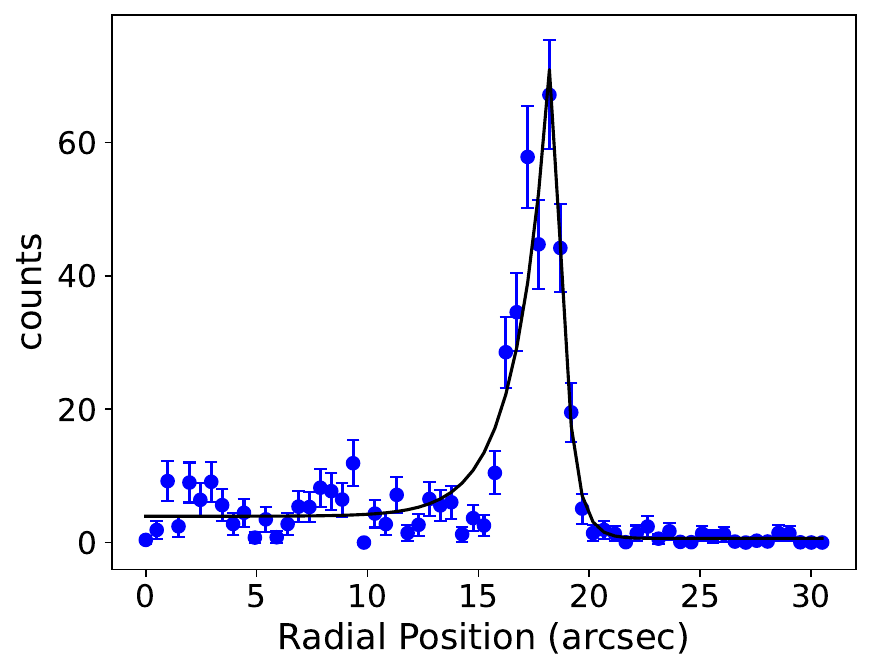}
  \end{subfigure}
  \caption{Filament profiles for some of the selected regions. Black lines indicate the best fit.}
  \label{line_profile}
\end{figure*}

The selected regions are shown in Fig.\ref{region_selection}. The width of the selected regions to measure the thickness was kept at about 3 arcsec, and for some cases it was kept at about 5 arcsec to ensure a large enough photon count. The region width was selected to ensure that we could measure the small-scale turbulence present in the system. 
The determined thickness of the filament profile in the radial direction is given in Table \ref{FWHM}. We exclude those regions for which the fitting using Eq.(\ref{eq2}) does not provide an acceptable fit with null hypothesis random probability larger than $0.1\%$ (for the current case with degree of freedom of 53, it corresponds to a reduced chi-square of $\chi_{\nu}^2=1.7$). 
The filament profiles for some of the regions are shown in Fig.\ref{line_profile}

\begin{table}
    \caption{Thickness of the X-ray rim (2.5 - 4 keV)}
    \label{FWHM}
    \begin{tabular}{lccr}
         \hline
         R.A ($^{\circ}$) & DEC ($^{\circ}$) & Thickness (arcsec) & $\chi^2_{\nu}$\\
         \hline
         6.1972042 & 64.0973531 & $3.21\pm0.21$ & 1.32\\
         6.1956346 & 64.098445 & $2.46\pm0.24$ & 1.2\\
         6.1940671 & 64.0991269 & $1.89\pm0.23$ & 1.13\\
         6.1928113 & 64.0999458 & $1.71\pm0.22$ & 0.7\\
         6.1912408 & 64.1011742 & $2.05\pm0.21$ & 1.05\\
         6.1902987 & 64.1018567 & $2.09\pm0.14$ & 0.79\\
         6.1890437 & 64.1025389 & $1.78\pm0.23$ & 1.03\\
         6.1877879 & 64.1033578 & $1.3\pm0.16$ & 1.38\\
         6.1868442 & 64.1043133 & $2.35\pm0.33$ & 1.1\\
         6.1814937 & 64.1100481 & $1.69\pm0.13$ & 1.47\\
         6.1802362 & 64.11114 & $1.67\pm0.2$ & 0.96\\
         6.1796055 & 64.1119595 & $1.84\pm0.18$ & 0.91\\
         6.1786618 & 64.1127785 & $1.44\pm0.12$ & 1.16\\
         6.1777185 & 64.1135975 & $2.58\pm0.36$ & 0.68 \\ 
         6.1767742 & 64.1145531 & $1.56\pm0.17$ & 0.7\\
         6.1761425 & 64.1155092 & $1.19\pm0.16$ & 1.02\\
         6.1755117 & 64.116465 & $1.12\pm0.13$ & 1.24\\
         6.1748805 & 64.1172844 & $1.2\pm0.25$ & 0.99\\
         6.1739358 & 64.11824 & $1.12\pm0.1$ & 0.93\\
         6.1736192 & 64.1189233 & $1.15\pm0.14$ & 1.01\\
         6.1729876 & 64.1198792 & $0.9\pm0.1$ & 1.19\\
         6.1726692 & 64.1208356 & $1.18\pm0.12$ & 0.79\\
         6.1720371 & 64.1217914 & $1.16\pm0.18$ & 1.08\\
         6.1714054 & 64.1227472 & $1.06\pm0.15$ & 1.15\\
         6.1710879 & 64.1235672 & $1.24\pm0.18$ & 1.06\\
         6.170457 & 64.1243863 & $1.6\pm0.25$ & 0.77\\
         6.1698254 & 64.1253425 & $1.05\pm0.13$ & 0.8\\
         6.1691933 & 64.1262986 & $2.01\pm0.24$ & 0.65\\
         6.1688746 & 64.1272544 & $1.58\pm0.14$ & 1.1\\
         6.1682425 & 64.1282106 & $1.02\pm0.09$ & 0.77\\
         6.1685496 & 64.1293044 & $1.47\pm0.15$ & 0.93\\
         6.1672918 & 64.1301228 & $1.08\pm0.13$ & 0.4\\
         6.1685398 & 64.1309444 & $1.06\pm0.15$ & 1.04\\
         6.1685344 & 64.131901 & $1.06\pm0.15$ & 0.68\\
         6.1685287 & 64.1328578 & $1.27\pm0.21$ & 0.7\\
         6.1706692 & 64.1420167 & $1.71\pm0.26$ & 1.16\\
         6.1706612 & 64.1433833 & $0.97\pm0.13$ & 1.59\\
         6.1706533 & 64.14475 & $2.2\pm0.32$ & 1.27\\
         6.1725271 & 64.1459822 & $1.44\pm0.11$ & 1.48\\
         6.173149 & 64.1468026 & $1.38\pm0.12$ & 1.32\\
         6.1734573 & 64.1477598 & $1.22\pm0.09$ & 1.52\\
         6.1737659 & 64.14858 & $1.32\pm0.1$ & 1.37\\
         6.1747008 & 64.1495375 & $1.11\pm0.09$& 1.78\\
         6.1746955 & 64.1504945 & $1.1\pm0.16$ & 1.3\\
         6.1753171 & 64.1514517 & $1.28\pm0.16$ & 1.3\\
         6.1762521 & 64.1524092 & $1.41\pm0.22$ & 0.93\\
         6.1806004 & 64.1599306 & $2.61\pm0.42$ & 1.83\\
         6.1821588 & 64.1617086 & $1.75\pm0.25$ & 1.43\\
         6.1834063 & 64.1629398 & $1.6\pm0.19$ & 1.07\\
         6.1846554 & 64.1640347 & $2.53\pm0.18$ & 1.77\\
         6.1871526 & 64.1662239 & $4.31\pm0.87$ & 1.73\\
         6.2121854 & 64.1797761 & $1.66\pm0.16$ & 1.14\\
         6.2137504 & 64.1807339 & $1.38\pm0.11$ & 1.36\\
         6.2153158 & 64.1815553 & $1.51\pm0.12$ & 0.73\\
         6.2178246 & 64.1819672 & $1.12\pm0.07$ & 1.15\\
         6.2193913 & 64.182515 & $0.85\pm0.06$ & 1.01\\
         6.2206437 & 64.183336 & $0.82\pm0.06$ & 0.82\\
         6.2228383 & 64.1838842 & $0.91\pm0.06$ & 0.63\\
         6.2247192 & 64.1844322 & $1.31\pm0.09$ & 1.4\\
         6.2265988 & 64.1852536 & $2.34\pm0.41$ & 1.35\\
         6.347444 & 64.2097523 & $1.39\pm0.25$ & 1.06\\
         6.3496428 & 64.2097517 & $1.29\pm0.3$ &0.7\\
         6.3672287 & 64.206195 & $1.68\pm0.27$ & 0.66\\

    \end{tabular}
\end{table}
\begin{table}
    \begin{tabular}{lccr}
         \hline
         R.A & DEC & FWHM (arcsec) & $\chi^2_{\nu}$\\
         \hline
         6.3694269 & 64.2059212 & $1.44\pm0.16$ & 0.67\\
         6.371625 & 64.2055106 & $1.21\pm0.12$ & 1.31\\
         6.3738229 & 64.2053733 & $1.86\pm0.24$ & 0.76\\
         6.3757059 & 64.204416 & $2.91\pm0.37$ & 1.35\\
         6.3775892 & 64.2040055 & $2.79\pm0.32$ & 1.31\\
         6.4932746 & 64.1167283 & $2.12\pm0.48$ & 0.85\\
         6.4926438 & 64.115909 & $2.06\pm0.28$ & 1.2\\
         6.4901361 & 64.115365 & $1.74\pm0.26$ & 0.99\\
         6.4888779 & 64.1142728 & $3.69\pm0.67$ & 1.17\\
         6.4788092 & 64.1047172 & $1.78\pm0.25$ & 1.06\\
         6.4775527 & 64.1038982 & $1.45\pm0.14$ & 1.44\\
         6.4762963 & 64.1029428 & $2.36\pm0.24$ & 1.3\\
         6.4750396 & 64.1019875 & $2.33\pm0.3$ & 1.1\\
         6.4715596 & 64.1558376 & $2.3\pm0.26$ & 1.52\\
         6.4728008 & 64.1533764 & $2.16\pm0.13$ & 1.45\\
         6.4734235 & 64.1525559 & $2.21\pm0.11$ & 1.17\\
         6.4740461 & 64.1517353 & $3.06\pm0.19$ & 1.55\\
         6.4771445 & 64.1448988 & $2.46\pm0.19$ & 1.13\\
         6.2451246 & 64.0793489 & $1.16\pm0.19$ & 0.77\\
         6.2423079 & 64.0803039 & $2.07\pm0.38$ & 1\\
         6.3101637 & 64.0681656 & $2.37\pm0.25$ & 0.86
    \end{tabular}
\end{table}

\begin{figure}
    \includegraphics[width=0.45\textwidth]{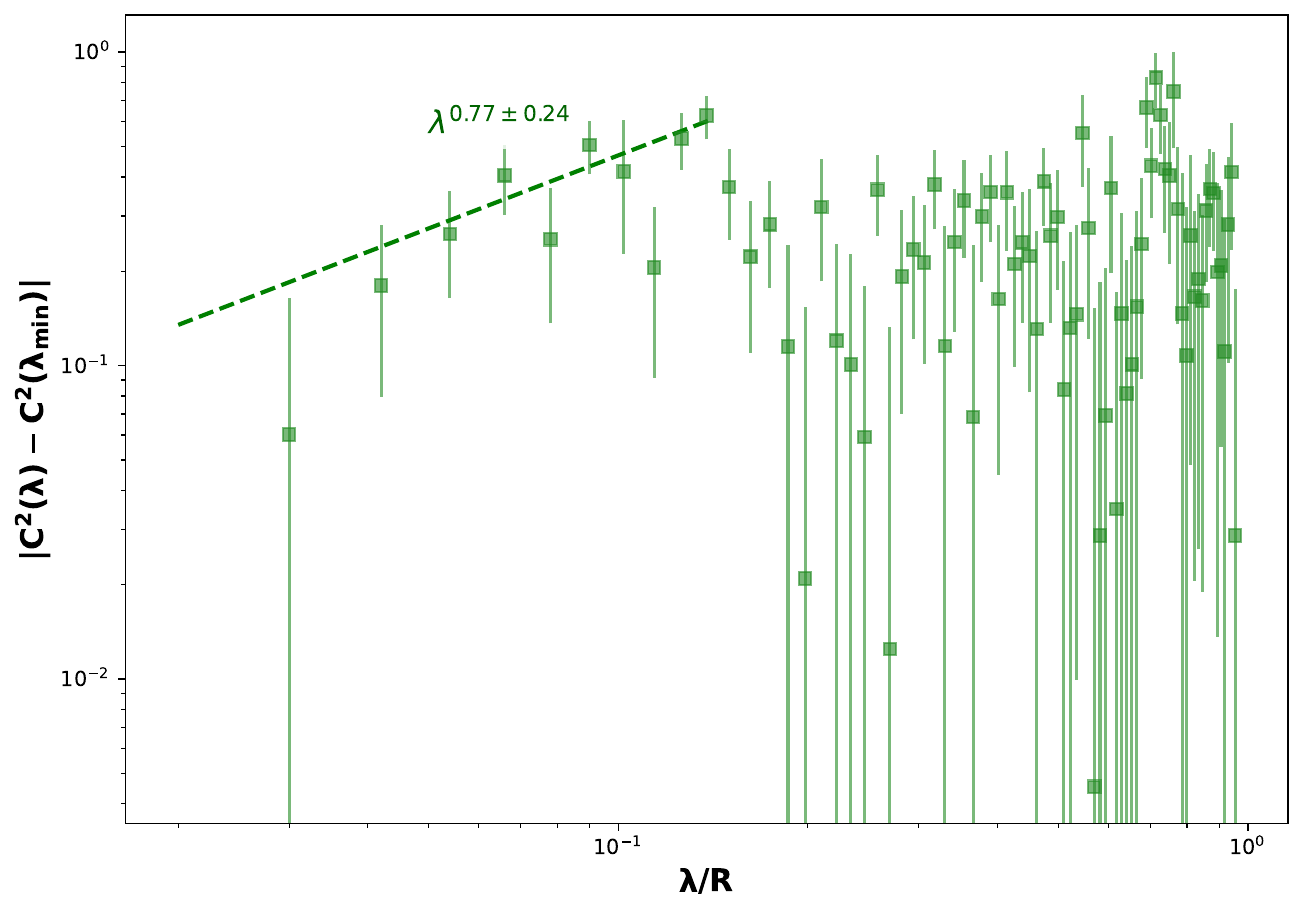}
    \caption{Magnetic energy spectrum obtained using the X-ray rim thickness.}
    \label{two-point_fwhm}
\end{figure}

\begin{figure}
    \includegraphics[width=0.45\textwidth]{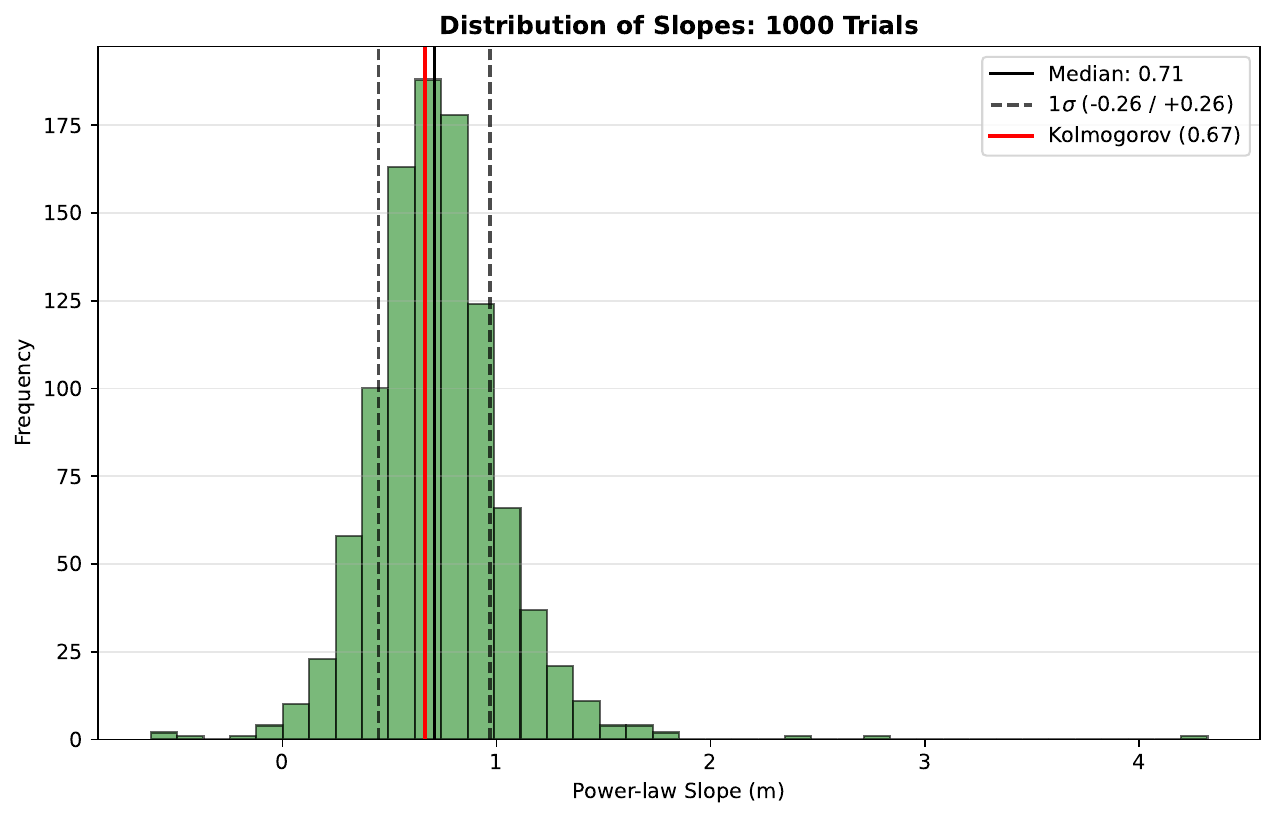}
    \caption{MC simulations for the slope of two-point correlation obtained using X-ray rim thickness}
    \label{two-point_MC}
\end{figure}

To remove the spatial dependency of the density of electrons in the measured turbulence, we use the FWHM of the X-ray rim from Table \ref{FWHM} in the two-point correlation. This result is plotted in Fig. \ref{two-point_fwhm}. The data points are associated with large uncertainity, but we could see that the turbulence is developed up to $\lambda/r \leq0.1$ with a power law scaling of $\lambda^{0.77\pm0.24}$. 
We performed Monte Carlo simulations to test the robustness of the above results by propagating the measurement uncertainties. The slope was fitted over 1000 trials, and the resulting distribution is shown in Fig.\ref{two-point_MC}. The simulations yield a median slope of 0.71 with a $1\sigma$ uncertainty of $\pm0.26$.
This result is consistent with the results obtained using the X-ray flux image for the region $r\geq0.9$ in Tycho.

\section{Discussion}
The correlation function of the X-ray image can probe the magnetic turbulence in the immediate acceleration sites where high-energy electrons are accelerated, since these short-lived electrons ($<$100 years) trace the turbulence present in these plasmas. In contrast, radio-emitting electrons, due to their longer lifetimes, sample plasmas across a broader range of regions \citep[][]{2018Galax...6...46V}. The correlation for $r\geq0.9 R_{\rm SNR}$  from radio synchrotron image showed that the turbulence scale was developed up to $\lambda/r\leq0.5$ \citep[][]{2018MNRAS.480.2200S}, while the turbulence scale measured with X-ray flux image in this study showed that the magnetic turbulence was developed up to $\lambda/r\leq0.1$ for $r\geq0.88 R_{\rm SNR}$.   

A similar result was reported in \citet{2021MNRAS.502.5313S}, in which an auto-angular power spectrum is used to measure the magnetic turbulence using radio and X-ray image (4.2-6.0keV) of Cas A with results indicating a smaller angular scale in X-ray than in radio. 
This different turbulence scale is also consistent with the radio and X-ray polarization results of Tycho's SNR. The polarization studies of Tycho at the rim closer to the shock region indicated a polarization degree of 7-8$\%$ \citep[][]{1971ApJ...163..231K, 1991AJ....101.2151D} for the radio band and a higher polarization degree of $12\%\pm2\%$ for the X-ray band \citep[][]{2023ApJ...945...52F}. 
The observed smaller scale turbulence from the X-ray flux image may also be due to the shorter lifetime of high-energy electrons emitting X-ray photons, because of their shorter lifetime they cannot travel further and might trace the small-scale turbulence in the active acceleration site, unlike electrons emitting radio photons, which have longer lifetime, so they can travel further from the post-shock sites. The magnetic field in these two regions can be quite different and thus resulting in different polarization degrees \citep[][]{2025ApJ...986..210Z} .

A power-law scaling with power index 2/3 was observed in the magnetic energy spectrum derived from radio synchrotron images of Tycho \citep[][]{2018MNRAS.480.2200S} and Cas A \citep[][]{2020MNRAS.498.1093V}, suggesting the presence of Kolmogorov-like turbulence in these systems. \citet{2025ApJ...994..147P} applied a similiar two point correlation along the concentric circle for thermal X-ray image of Tycho's SNR in the energy range of 1.2-4 keV, this spectrum traces the density fluctuations and they noted that these follow a 2/3 law except for the region r=0.9$R_{\rm SNR}$ and r=0.95$R_{\rm SNR}$.
In contrast, the magnetic energy spectrum obtained from X-ray flux for Tycho's SNR shows a significantly steeper slope for region $r \geq 0.9 R_{\rm SNR}$ as shown in Table \ref{pl_twopoint}.
On the other hand, the spectrum derived from the X-ray rim thickness is independent of electron number density and reveals a turbulence scale extending up to $\lambda/r \leq 0.1$, with a power-law slope of $0.77 \pm 0.24$, as shown in Fig.~\ref{two-point_fwhm}. The steeper slope observed in the spectrum measured from the X-ray flux image may be influenced by the contribution of electron number density. The 2/3 power law is observed in the magnetic energy spectrum derived from the X-ray rim thickness is broadly consistent with a Kolmogorov-like turbulence spectrum within the uncertainties. We note that the correlation function becomes quite chaotic for the region of $r < 0.90 R_{\rm SNR}$. 
The observed  turbulence scale cannot be explained by the Bell instability \citep[][]{2004MNRAS.353..550B} caused by the streaming of cosmic ray particles, as the maximum turbulence scale caused by this effect in amplified magnetic field of $100\mu G$ by knee-energy cosmic rays would be
$\mathrm{\lambda_{max}\approx10^{16}cm}$,  which is much smaller than the turbulence scale of 25.1 arcseconds ($\approx10^{18}$cm) that we measured. The observed turbulence scale is unlikely to have been generated by cosmic-ray streaming, suggesting that it was already present before the arrival of the SNR shock.

While the two-point correlation analysis provides a robust measure of spatial structure, we note that the features probed near the shock front may reflect the turbulence scale in the upstream region. As discussed by \citet{2008ApJ...678..939Z} and \citet{2025ApJS..277...44P}, pre-existing density or magnetic-field inhomogeneities in the upstream medium can be altered by the shock front, imprinting their structure on the downstream emission. These structures could potentially be misinterpreted as signatures of a turbulence cascade.
\citet{2025Univ...11..377S} proposed that Tycho's supernova exploded within a planetary nebula. If that is the case, the spatial scale measured by the two-point correlation in the downstream region could partly reflect pre-existing structures in the circumstellar medium that were present before the passage of the SNR shock.

\section*{Acknowledgements}

This research has made use of data obtained from the Chandra Data
Archive, and software provided
by the Chandra X-ray Center in the application packages CIAO. This research has made use of data
and software provided by the High Energy Astrophysics Science
Archive Research Center, which is a service of the Astrophysics
Science Division at National Aeronautics and Space Administration/Goddard Space Flight Center . This work was supported by the National Science and Technology Council (NSTC) of the Republic of China (Taiwan) under grants NSTC 113-2112-M-007-020 and 114-2112-M-007-042, Taiwan Space Agency (TASA) under grants
NSPO-P-109221 and TASA-P-1140072, the National SKA Program of China (2025SKA0140100), the National Key R\&D Program of China (No. 2023YFA1608000), the China Manned Space Program with grant no. CMS-CSST-2025-A14, and the Youth Innovation Promotion Association of CAS (2023000015).

\section*{Data Availability}
The data used for the analysis and findings within this paper will be
shared by the corresponding author upon reasonable request.



\bibliographystyle{mnras}
\bibliography{example} 

\appendix
\section{}
\begin{figure*}
  \begin{subfigure}{\columnwidth}
  \includegraphics[width=\textwidth]{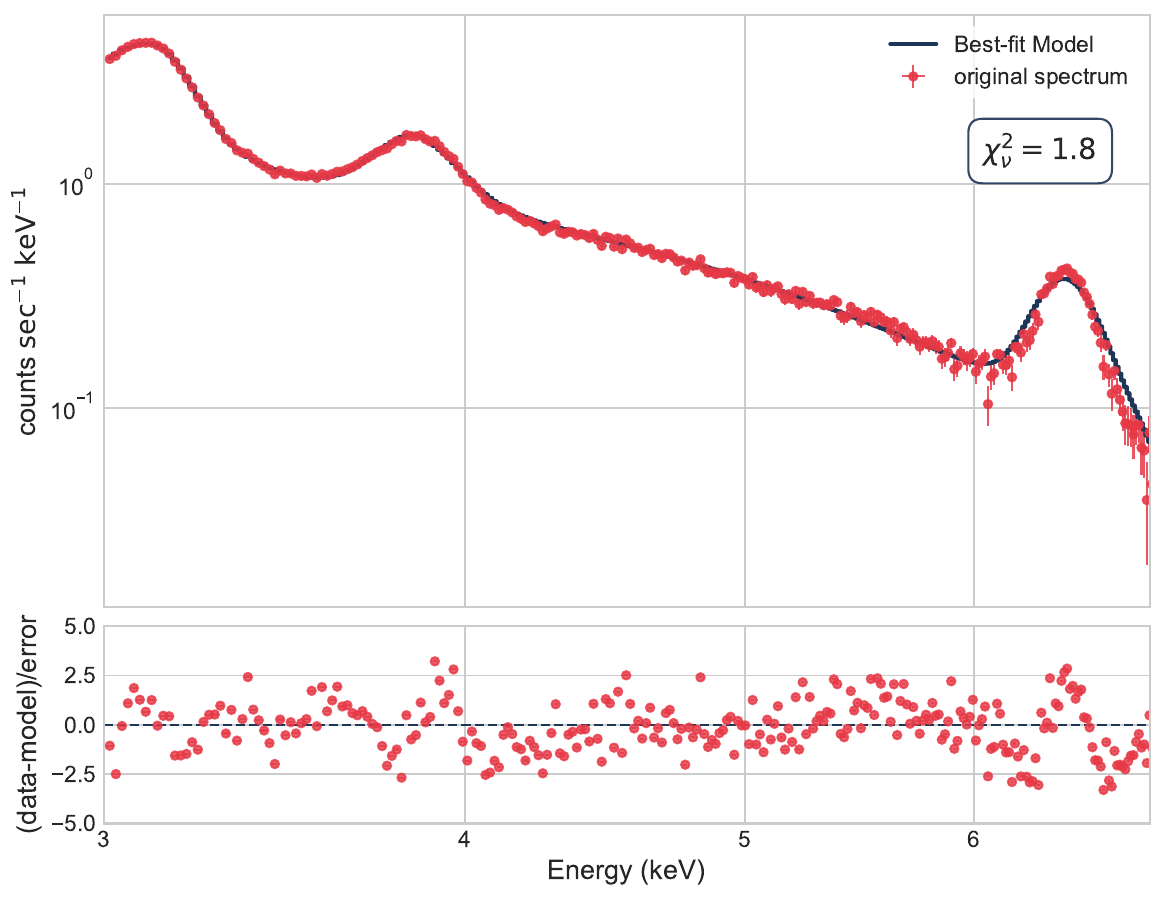}
  \end{subfigure}
  \hfill
  \begin{subfigure}{\columnwidth}
  \includegraphics[width=\textwidth]{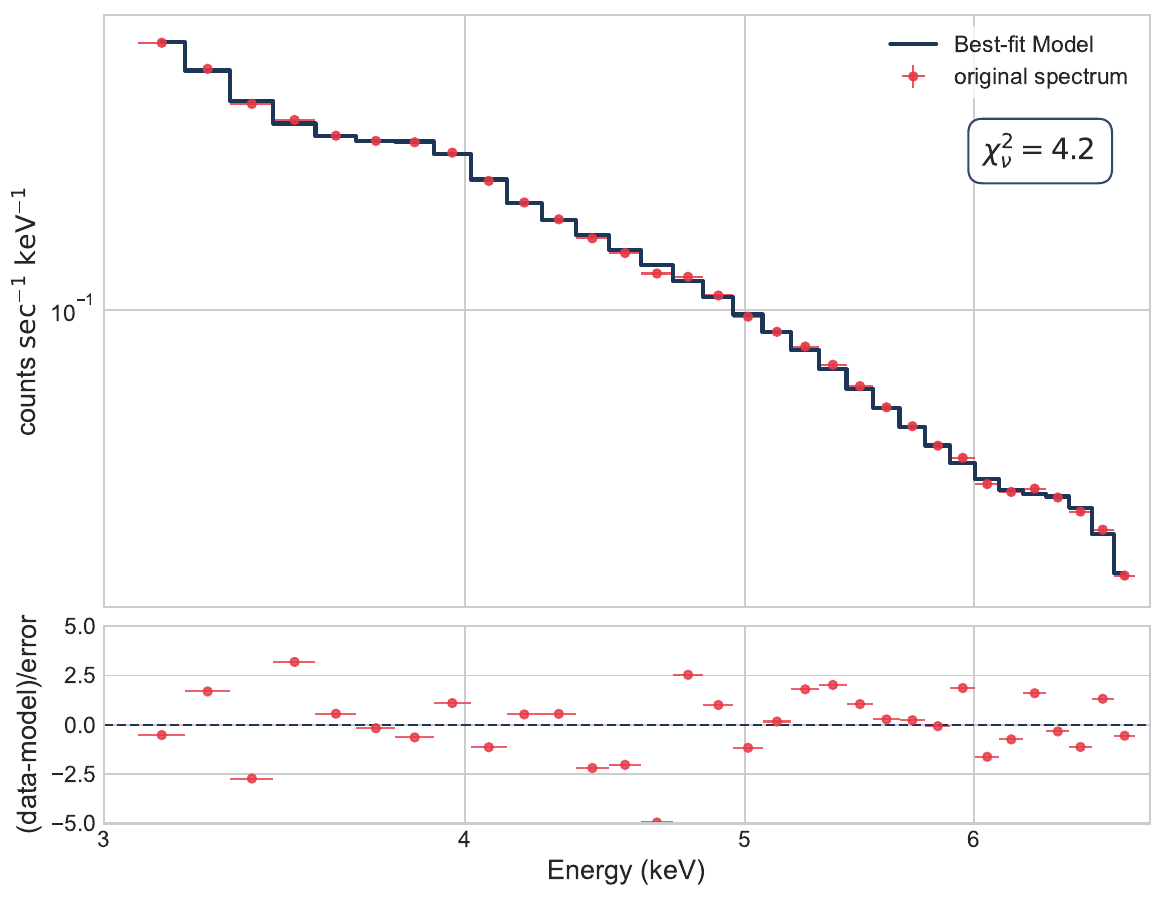} 
  \end{subfigure} 
  \caption{Folded model fitted with a power-law + three gaussian model is shown. Right panel shows the original spectrum and the left panel shows the pGMCA extracted spectrum}
  \label{spectrum-reference}
\end{figure*}
As shown in Fig. \ref{fig:energy_Count}, the original spectrum contains a substantial thermal contribution, while residual thermal emission is still present in the pGMCA-extracted spectrum. To quantify the thermal contamination, the spectra were fitted in XSPEC using a power-law component to model the non-thermal emission and three Gaussian components to represent the thermal line features. Both original and pGMCA-extracted spectra were fitted using the same model and is shown in Fig. \ref{spectrum-reference}. The combined thermal flux ($F_{\rm thermal}$) from the three Gaussian components and the non-thermal flux ($F_{\rm non\mbox{-}thermal}$) from the power-law component were then calculated for each spectrum. We find that the ratio $F_{\rm thermal}/F_{\rm non\mbox{-}thermal}$ decreases from $43\%$ in the original spectrum to $8\%$ in the pGMCA-extracted spectrum.





\bsp	
\label{lastpage}
\end{document}